\documentclass[fleqn,usenatbib]{rasti}

\usepackage{newtxtext,newtxmath}
\usepackage[T1]{fontenc}

\DeclareRobustCommand{\VAN}[3]{#2}
\let\VANthebibliography\thebibliography
\def\thebibliography{\DeclareRobustCommand{\VAN}[3]{##3}\VANthebibliography}

\usepackage{graphicx}	
\usepackage{amsmath}	
\usepackage{orcidlink}
\usepackage{booktabs}
\usepackage{lineno}

\title[TESSExtractor: TESS FFI Light-Curves]{TESSExtractor: A web-based interactive tool for light-curve visualisation and period estimation using TESS full-frame images.}

\author[J. Serna et al.]
{Javier Serna$^{1}$\thanks{E-mail: jserna@ou.edu}\orcidlink{0000-0001-7351-6540},
Jesús Hernández$^{2}$\orcidlink{0000-0001-9797-5661},
Sean P. Matt$^{1}$\orcidlink{0000-0001-9590-2274},
Giovanni Pinzón$^{3}$\orcidlink{0000-0001-9147-3345},
Astaroth Elizabethson$^{4}$\orcidlink{0000-0002-2249-8474},
\newauthor
Jorge Martínez-Palomera$^{5,6}$\orcidlink{0000-0002-7395-4935},
Carlos G. Román-Zúñiga$^{2}$\orcidlink{0000-0001-8600-4798},
Sergio Sánchez-Sanjuán$^{2}$\orcidlink{0000-0002-2269-9348},
Jaime Morice$^{1}$\orcidlink{0009-0003-0564-9423},
\newauthor
Thanawuth Thanathibodee$^{7}$\orcidlink{0000-0003-4507-1710},
Margarita Pereyra$^{1,2,15}$\orcidlink{0000-0001-6148-6532},
Mauricio Tapia$^{2}$\orcidlink{0000-0002-0506-9854},
Alejandro García-Varela$^{4}$\orcidlink{0000-0001-8351-0628},
\newauthor
Juan F. Cabrera-García$^{4,8}$\orcidlink{0000-0002-0020-843X},
Marcelo Jaque Arancibia$^{9}$\orcidlink{0000-0002-8086-5746},
María Gracia Batista$^{3}$\orcidlink{0009-0009-3333-9477},
Ricardo López-Valdivia$^{2}$\orcidlink{0000-0002-7795-0018},
\newauthor
Raúl Michel$^{2}$\orcidlink{0000-0003-1263-808X},
Ezequiel Manzo-Martínez$^{12}$\orcidlink{0000-0001-6647-862X},
Keivan G. Stassun$^{10}$\orcidlink{0000-0002-5365-1267},
Cesar Briceño$^{11}$\orcidlink{0000-0001-7124-4094},
Sydney Robertson$^{1}$,
\newauthor
Tanay Ayiliath$^{1}$\orcidlink{0009-0005-4471-208X},
David Fernández$^{9}$\orcidlink{0009-0002-6359-7513},
Fabián Ahumada$^{13}$,
Alejandro Alfonso$^{14}$\orcidlink{0009-0001-5527-483X}
\\
$^{1}$Homer L. Dodge Department of Physics and Astronomy, University of Oklahoma, Norman, OK 73019, USA\\
$^{2}$Universidad Nacional Autónoma de México. Instituto de Astronomía. A.P. 106, 22800. Ensenada, B.C. , México\\
$^{3}$Observatorio Astron\'omico Nacional, Facultad de Ciencias, Universidad Nacional de Colombia, Bogot\'a, Colombia\\
$^{4}$Universidad de los Andes, Departamento de F\'{\i}sica, Cra. 1 No. 18A-10, Bloque Ip, AA 4976, Bogot\'a, Colombia\\
$^{5}$NASA Goddard Space Flight Center, 8800 Greenbelt Road, Greenbelt, MD 20771, USA\\
$^{6}$University of Maryland, Baltimore County, 1000 Hilltop Circle, Baltimore, MD 21250, USA\\
$^{7}$Department of Physics, Faculty of Science, Chulalongkorn University, 254 Phayathai Road, Pathumwan, Bangkok 10330, Thailand\\
$^{8}$Dipartimento di Fisica e Astronomia ``Galileo Galilei", Università degli Studi di Padova, Vicolo dell’Osservatorio 3, I-35122 Padova, Italy\\
$^{9}$Departamento de Astronom{\'i}a, Universidad de La Serena, 1700000 La Serena, Chile\\
$^{10}$Department of Physics and Astronomy, Vanderbilt University, Nashville, TN 37235, USA\\
$^{11}$Cerro Tololo Inter-American Observatory, Casilla 603, La Serena, Chile\\
$^{12}$Department of Astronomy, University of Michigan, Ann Arbor, MI 48109, USA\\
$^{13}$Departamento de F\'{\i}sica, Universidad de Concepci\'on, Concepci\'on, Chile\\
$^{14}$Universidad Pedagógica y Tecnológica de Colombia, Tunja, Colombia\\
$^{15}$Secretar\'ia de Ciencia, Humanidades, Tecnolog\'ia e Innovaci\'on, Research Fellow\\
}

\date{Accepted XXX. Received YYY; in original form ZZZ}

\pubyear{\the\year{}}

\begin{document}
\label{firstpage}
\pagerange{\pageref{firstpage}--\pageref{lastpage}}
\maketitle

\begin{abstract}
We present a user-friendly web application that allows users to extract, visualise, and interact with TESS light-curves (LCs) from full-frame images. The service works for any star identified in the TIC database or using stellar coordinates. TESSExtractor performs photometry using a circular aperture centred on the object of interest, with a sky annulus for background subtraction. The application also uses cotrending basis vectors provided by TESS to correct photometry for possible systematic effects.
The Lomb-Scargle Periodogram is used to obtain periods from the LC. The TESSExtractor's website offers an intuitive interface for scientific and educational purposes and is publicly available at \url{https://www.tessextractor.app}. This tool simplifies the analysis of TESS LCs and provides valuable insight for researchers seeking to extract and analyse TESS data for a variety of scientific goals (e.g., stellar rotation, activity, transient detection, asteroseismology, or simply detection of planetary transits or stellar companions, among others). TESS provides insights into our understanding of time-domain astrophysics, and TESSExtractor provides an accessible and intuitive web interface for scientific exploitation of TESS data. We demonstrate that TESSExtractor produces LCs of sufficient quality for both quantitative and morphological time-domain studies.
\end{abstract}




\section{Introduction}

The Transiting Exoplanet Survey Satellite (TESS) mission was planned mainly to detect transiting exoplanets, yet it offers an exceptional opportunity to conduct variability studies that will allow us to better understand the astrophysical processes of the stars and planets. Currently, TESS has covered around 97\% of the entire sky and continues to expand its coverage through its extended missions, with around 27 days of continuous observations per sector. At present, TESS data provides an excellent opportunity for exoplanet science and constitutes a valuable dataset for studying phenomena related to brightness variability across a wide range of astrophysical contexts, beyond planetary science \citep{Ricker2014}.\\

\noindent In addition to planet detection and characterisation \citep[e.g.,][]{Newton2019, Rao2021, Melton2024}, TESS can detect brightness variability that can indicate stars with periodic phenomena, such as eclipses in binary systems \citep[e.g.,][]{prsa:2022} and rotational modulation caused by stellar spots \citep[e.g.,][]{Serna2021}. In particular, stellar spots are associated with stellar magnetic fields, and by studying their rotational modulation, we can obtain information about the internal structure, magnetic activity, and the evolution of stellar angular momentum \citep{Bouvier2014,Pinzon2021}. Measuring stellar brightness variations over time allows us to determine fundamental physical parameters, such as the stellar radius and spin axis inclination \citep[e.g.,][]{Kovacs2018,Bowler2023}, and even investigate how stars are affected by both internal processes and external factors, such as binarity and angular momentum transfer \citep{Kounkel_2021,Kounkel_2023}.\\

\noindent Each month, the Mikulski Archive for Space Telescopes (MAST) ingests and releases calibrated data products from the TESS mission. While MAST services pre-extracted light curves (LCs) for selected official target lists, such as those processed by the Science Processing Operations Center \citep[SPOC;][]{SPOC_2016,Caldwell_2020} and the Quick-Look Pipeline \citep[QLP;][]{Huang_2020} pipelines, it does not deliver LCs for every point source in the sky. The official target list includes a relatively limited set of targets selected from Guest Investigator programs, planet candidate lists, or curated priority catalogues. 
During its primary mission, TESS observed $\sim$200,000 pre-selected stars at 2-minute cadence \citep{Fetherolf_2023}, and community pipelines such as QLP extended this to $\sim$9.1 million sources brighter than $T_{\rm mag} = 13.5$ \citep{Kunimoto_2021}. However, the TESS Full-Frame Images (FFIs), currently collected at 200-second cadence, enabled the extraction of 
over 83 million light curves from Cycle 1 alone down to $T_{\rm mag} = 16$ \citep{Roth_2026}, nearly three orders of magnitude more than the official 2-minute cadence target list. This gap between mission-delivered products and the full detectable sky underscores the need for accessible, on-demand tools that operate directly on FFI data. For the vast majority of stars, only the raw FFIs from each sector are available, requiring users to perform their own photometric extraction and correction of systematic effects. As a result, researchers must rely on tools such as TESSCut \citep{TESSCut} to download pixel cutouts and perform aperture selection, background subtraction, quality-flag filtering, and optional cotrending corrections. While this workflow is powerful and flexible, it requires substantial coding skills, astrophysical expertise, and familiarity with the TESS data architecture. Critically, these extraction and correction steps are not consolidated into a general-purpose, community-wide automated pipeline accessible without programming expertise, creating a significant barrier for non-expert users and limiting reproducibility across studies.\\

\noindent Other significant efforts have also contributed to broadening community access to TESS data. For example, the QLP \citep{Huang_2020,Kunimoto_2021} has generated one of the largest databases of FFI-extracted LCs, although this analysis remains limited to stars brighter than $T_{\rm mag}=13.5$. In addition, several open-source packages have been developed for processing, analysing and visualising TESS LCs such as \texttt{Eleanor} \citep{Feinstein_2019}, \texttt{Lightkurve} \citep{Lightkurve_2018}, DIA \citep{Oelkers_2018}, \texttt{TGLC} \citep{Han_2023}, \texttt{TESSILATOR} \citep{Binks2024}, \texttt{CDIPS} \citep{Bouma_2019}, \texttt{TESSReduce} \citep{Ridden-Harper_2021}, \texttt{TASOC} \citep{Handberg_2021}.
These tools have been invaluable for enabling reproducible analyses and advancing specific scientific goals. Nevertheless, most require prior programming experience, mainly python, careful documentation handling, and, in some cases, the development of custom scripts, steps that can be time-consuming and challenging for non-expert users or those seeking rapid exploratory analyses.\\

\noindent TESSExtractor was designed to eliminate this technical barrier by automating the entire photometric extraction and correction sequence directly from the FFIs. The application performs the complete workflow, including TESSCut data retrieval, aperture photometry, background subtraction, and systematic correction in a single step, entirely online, through a web browser, and without requiring any local software installation or coding expertise.\\

\noindent In this work, we present the TESSExtractor web application as an accessible and interactive tool for astronomers, researchers, students, and amateur astronomers who wish to quickly and easily process TESS FFI data to explore variability and periodicity of any point-like or extended source listed in the TESS input catalogue \citep[TIC,][]{Stassun_2019}, or any position specified by coordinates. TESSExtractor simplifies the analysis and visualisation of the TESS LCs by allowing users to quickly access all relevant data products through a few simple steps. The users need to type the target name or coordinates into the search bar of the web application, and they will promptly have access to the LC plot, raw files, and plots to examine potential periodicities. Alternatively, users may submit a list of targets to be processed asynchronously through a queue system and once products are generated, they are automatically delivered to the user via email.\\

\noindent In Section~\ref{sec:Extraction} we present the methodology for extracting the LCs, the systematic effects correction, and the estimation of the contamination flag. Also, we provide a practical step-by-step guide for app users. In Section~\ref{sec:Applications}, we present the data products generated by TESSExtractor, validate the extracted LCs against those produced by other existing pipelines, and compare derived periods with results from independent analyses. Finally, in Section~\ref{sec:Discussion} we discuss the current advantages and limitations of TESSExtractor and outline future improvements planned for subsequent releases.

\section{Extracting LCs}
\label{sec:Extraction}
TESSExtractor can be accessed at \url{https://www.tessextractor.app}, where the users have three options for obtaining LCs. 1) {\it The default option} uses an automatically determined aperture and sky annulus that depends on the source brightness and extracts the LC without applying instrumental corrections. 2) {\it The systematic correction option} applies known trends to correct the LCs for potential instrumental systematics. In this mode, the users may also choose the aperture and sky annulus parameters for the LCs. 3) {\it The bulk search option} allows the users to submit a list of up to 100 targets; for each target, all
available TESS sectors are processed and the resulting data products, the corrected and uncorrected LCs, Lomb-Scargle periodograms, phase-folded LCs, data files, and a summary catalogue are delivered as a compressed archive via email.
In the following, we describe the LC extraction process in detail. 

\subsection{TESS Cutouts}
During its primary mission, TESS obtained FFIs for each sector at a 30-minute cadence. In its subsequent extended missions, the cadence was improved first to 10 minutes and later to 200 seconds, significantly enhancing the temporal resolution of the TESS observations. Each FFI contains on the order of $10^5-10^6$ detectable sources, depending on field density, with many sources brighter than $T_{mag} = 16$ which makes it challenging to extract photometry for the entire catalogue simultaneously.
The \texttt{TESScut} service \citep{TESSCut} provides an efficient solution by allowing users to request small subimages, or cutouts, centred on specific targets of interest. This approach substantially reduces the computational time and storage requirements needed to process large sequences of FFIs.\\

\noindent TESSExtractor downloads TESS pixel cutouts for each target using the \texttt{TESScut} service via the \texttt{Astropy} interface \citep{astropy:2013,astropy:2018,astropy:2022}. By default, this tool extracts a $10\times10$ pixel region from the FFI, centred on the object, although it also provides the option to retrieve larger regions of $15\times15$ and $20\times20$ pixels. The timestamps provided by \texttt{TESScut} are given in Barycentric TESS Julian Date (BTJD $=$ BJD$_{\rm TDB} - 2457000.0$), with the barycentric correction computed at the centre of the TESS CCD rather than at the precise sky position of each target. \texttt{TESSExtractor} does not apply any additional position-dependent time correction to account for the finite 
light-travel time across the focal plane or readout-related timing offsets. These corrections are typically of order 1--2\,s, negligible for the rotation and activity studies that constitute the primary use case of \texttt{TESSExtractor}, but users requiring precision timing (e.g., transit or eclipse timing variations) are advised to apply the appropriate corrections or to use pipeline products such as SPOC \citep{SPOC_2016} that incorporate them.

\begin{figure*}
    \centering
    \includegraphics[width=0.32\textwidth]{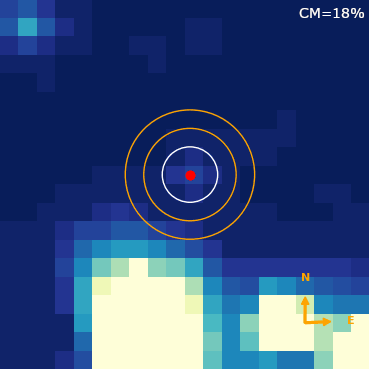}
    \includegraphics[width=0.325\textwidth]{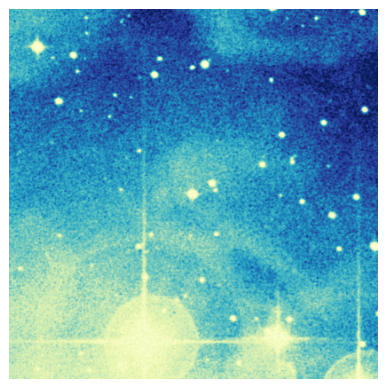}
    \includegraphics[width=0.32\textwidth]{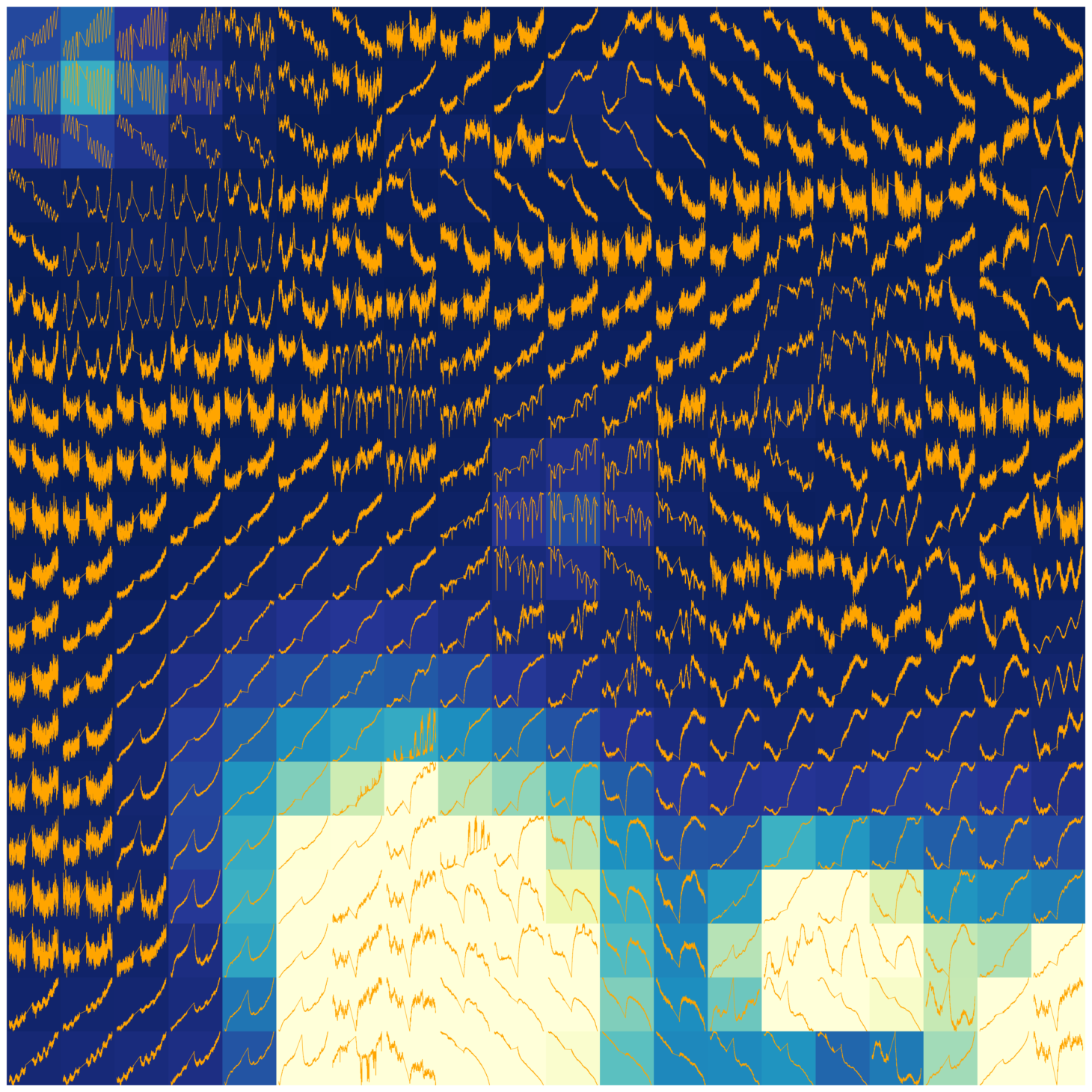}
    \caption{TESS, DSS2 and pixel-by-pixel diagnostics for the target Brun 691 (TIC 427393298) in Sector 6. Plots with the same Field of view (420 x 420 sq arcsec) correspond to a TESS image of 20x20 pixels. The white circle shows the photometric aperture, and the orange circle show the sky annulus. The right panel shows the pixel-by-pixel LCs across the TESS cutout, illustrating how the observed variability is spatially distributed within the field of view and enabling the identification of the pixels dominated by the target signal.
    } 
    \label{fig:cutout}
\end{figure*}

To support the visual inspection of the target in the TESS cutout image and to verify whether neighbouring sources contaminate the science source, TESSExtractor simultaneously downloads a Digital Sky Survey (DSS2 red) image with the same field of view as the TESS cutout\footnote{The DSS2 area is (420 arcsec)$^{2}$, equivalent to a (20$\times$20 pixel) TESS cutout.}, as shown in Figure \ref{fig:cutout}. The TESS image is rotated to match the standard astronomical orientation, with north up and east to the left, ensuring direct comparability with the DSS2 image.
Although the DSS2 Red band (6100–6900 Å) does not fully overlap with the broad TESS bandpass (6000–10000 Å), it provides a reasonably close approximation in the red optical regime, allowing a meaningful comparison of the sources detected in both datasets.

\subsection{Contamination Metric (CM)}
\label{sec:contamination}
\noindent Given the relatively large TESS pixel scale of 21 arcsec, the TESS data are highly susceptible to crowding, blending, and source confusion. In many cases, the LC may include flux contributions from nearby sources that fall within the chosen photometric aperture. To quantify the potential level of this contamination, we compute a \textit{contamination metric} defined as follows:

\begin{equation}
    CM=\left(1-\frac{f_{t}}{f_{t}+f_{n}}\right)\times100\,\,[\%]
\end{equation}

where $f_{t}$ is the flux of the target and $f_{n}$ is the total flux of Gaia sources that fall within the aperture radius, excluding $f_{t}$. The fluxes are based on the Gaia $G$-band photometry from Gaia-DR2 \citep{Gaia_2018}.
This metric represents the fraction of the total flux within the aperture that originates from neighbouring sources, and therefore provides an estimate of the potential level of blending rather than a direct measurement of photometric contamination. High CM values indicate fields in which flux dilution may significantly affect the extracted photometry. Following \citet{Handberg_2021}, we recommend treating targets with $CM > 15\%$ with caution, as their LCs are likely affected by crowding. 
The computed CM value is displayed in the web portal immediately after downloading each TESS cutout, allowing users to interactively adjust the aperture size and evaluate changes in this parameter. However, in very dense fields, even reducing the aperture to its minimum size may not significantly improve the CM, and users should therefore exercise caution when interpreting these LCs. We note that the CM is an approximation, it relies on Gaia~DR2 $G$-band photometry and a fixed circular aperture, and does not account for the actual TESS Pixel Response Function (PRF) or its sector-to-sector variations. Mission-derived alternatives such as the contamination ratio (\texttt{ContamRatio}) in the TIC~v8.2 \citep{Stassun_2019} or the \texttt{CROWDSAP} keyword in SPOC light curve products \citep{Caldwell_2020} use PRF-weighted flux estimates and may therefore differ from the CM values reported here, particularly in crowded fields. A formal comparison between these metrics is deferred to future work.

\subsection{Aperture selection and photometry}
\label{sec:photometry}
After downloading the TESS cutout, our pipeline uses the \texttt{Photutils} package in Python \citep{Bradley_2019} to perform simple aperture photometry (SAP). In the default and bulk options, the aperture for the target is determined based on its TESS magnitude ($T_{\rm mag}$) as catalogued in \citet[][TIC v8.2]{Stassun_2019}. Table \ref{tab:aperture_radius} summarises the default aperture radii and sky-annulus ranges adopted as a function of $T_{\rm mag}$, which can be overridden by the user in the photometry settings panel.

\begin{table}
\centering
\caption{Default aperture radii and sky-annulus ranges, also in pixels, as a function of target TESS magnitude $T_\mathrm{mag}$, adopted from TIC v8.2 \citep{Stassun_2019}. Users can override these values via the \emph{Photometry Settings} panel. Aperture radii are given in pixels; sky-annulus ranges are given as inner–outer radii.}
\label{tab:aperture_radius}
\begin{tabular}{ccc} 
 \hline
 $T_{mag}$ & Aperture radius (px) & Sky Annulus (px) \\
 \hline\hline
 $\leq 9$ & 3.0 & 4.0-5.0 \\
 $> 9$, $\leq 11$ & 2.5 & 3.5-4.5 \\
 $> 11$, $\leq 13$ & 2.0 & 3.0-4.0 \\
 $> 13$ & 1.5 & 2.5-3.5 \\
 \hline
\end{tabular}
\end{table}

The background subtraction of the photometry is performed using the mode of the pixel fluxes within the sky annulus. 
We remove all data points with bad TESS quality flags from the raw data to avoid anomalies in the photometry (e.g., cosmic rays, popcorn noise, and fireworks)\footnote{TESS Data release notes: \url{https://archive.stsci.edu/tess/tess_drn.html}}. 
In addition, we mitigate contamination from scattering light patterns on the TESS detector by rejecting all data points for which the sky-annulus flux exceeds the 95th percentile of its distribution.\\

The transformation of raw flux to magnitude for the LCs is computed as follows:
\begin{equation}
\label{eq:mag}
    m=-2.5\log{F_{\ast}}+ZP
\end{equation}

\noindent where $F_{\ast}$ is the flux after subtracting the background. ZP is the zero point of the TESS filter band. Despite the FFI's cadence changes between the prime and extended TESS missions, we adopt an average zero-point (ZP) value of 20.44 $\pm\,0.05$ \citep{Fausnaugh_2021}. Following standard error propagation, the uncertainty of $m$ is computed as:
\begin{equation}
\label{eq:magerr}
    \delta m = \sqrt{\left( 1.085 \frac{\delta F}{F_{\ast}} \right)^2 + (\delta ZP)^2}
\end{equation}
where $\delta F$ is the uncertainty in the counts obtained from the TESS cutout, and $\delta ZP$ is the uncertainty in the zero-point.\\

The default and the bulk options give the LCs calibrated to the $T_{mag}$ reported for the star in \citet[][TIC v8.2]{Stassun_2019}. 
We make an aperture correction of the LC magnitude using the $m$ time series, its median value ($\overline{m}$), and $T_{\rm mag}$. Using the following two-part function, we calibrate the LC magnitude to the reported TESS magnitude.

\[ m^{\ast} =
\left\{
\begin{array}{ll}
     m-|T_{\rm mag}-\overline{m}|,\,\,\,\,\,\,\,\, & \overline{m}>T_{\rm mag},\\
     m+|T_{\rm mag}-\overline{m}|,\,\,\,\,\,\,\,\, & \overline{m}<T_{\rm mag}\\
\end{array} 
\right. \]

It is important to note that $m^{\ast}$ is not an absolute photometric measurement but a relative calibration that anchors the median of the observed LC to the catalogue value $T_{\rm mag}$. This correction implicitly assumes that the median instrumental magnitude $\overline{m}$ during the observed sector is representative of the brightness at the epoch when $T_{\rm mag}$ was determined. For sources with genuine astrophysical variability such as eruptive young stellar objects, long-period variables, or eclipsing binaries with significant duty cycles, the TIC $T_{\rm mag}$ may not represent the true mean brightness during the observed sector, and the reported $m^{\ast}$ values should therefore not be interpreted as calibrated magnitudes.

\begin{figure}
    \centering
    \includegraphics[width=0.5\textwidth]{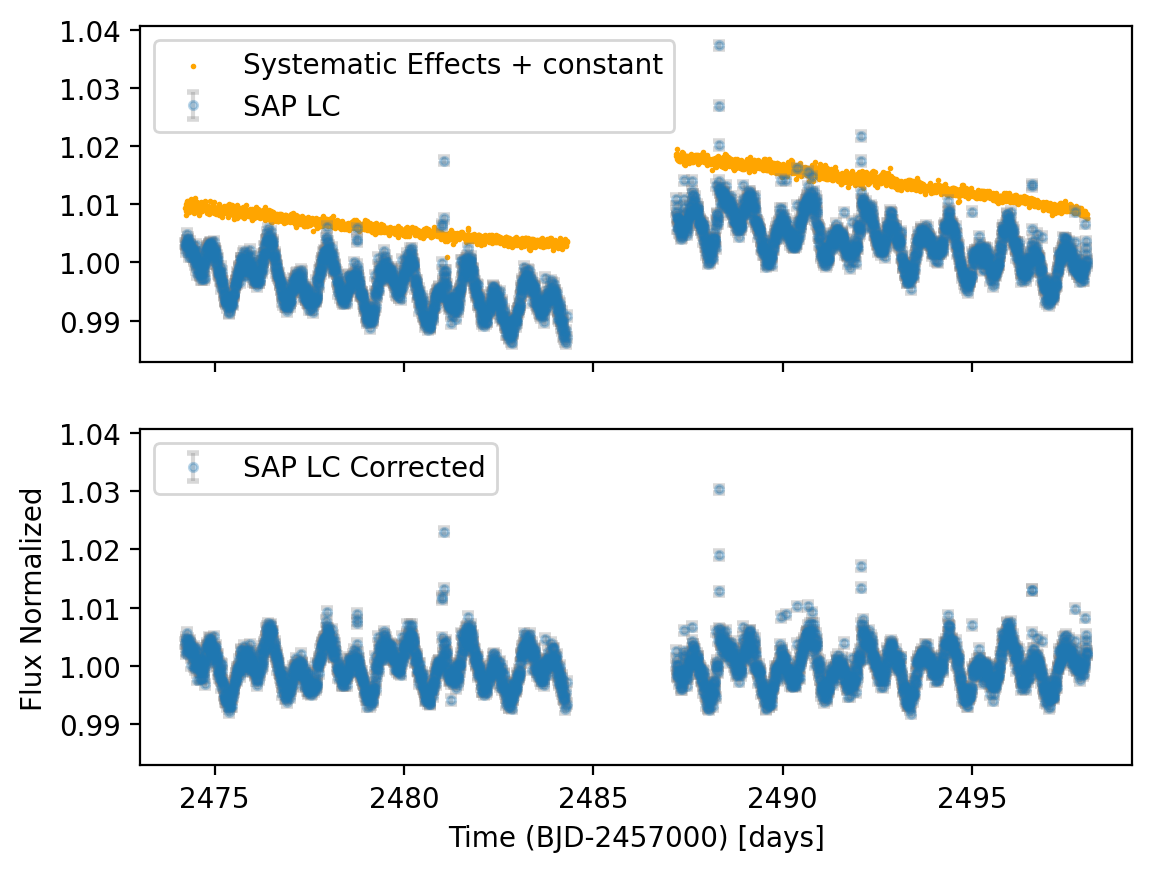}
    \caption{Systematic correction for the TESS LC TIC 60874953. In the top panel, we plot the best fit of the first four CBVs (\S \ref{sec:systematics}) to the normalised LC. In the lower panel, we plot the normalised LC after the systematic trend correction. 
    } 
    \label{fig:sys_corr}
\end{figure}

\subsection{Frame-by-Frame Cutout Animation}
\label{sec:frame_by_frame}
To visualise the spatial and temporal evolution of the flux within the TESS aperture, the advanced option of TESSExtractor generates a frame-by-frame animated sequence of the calibrated cutouts for each target.
Each frame corresponds to an individual cadence in the FFI sequence and displays the native TESS pixel array without reprojection. The position of 
the photometric aperture and sky annulus overlays is computed once from the sector-level WCS solution stored in the FITS aperture extension of the 
TESScut cutout, and remains fixed in pixel coordinates throughout the animation. The animation allows users to track temporal variations in brightness for both the science target and neighbouring sources, revealing pointing jitter, background variability, and scattered-light features that could affect the photometry.

The central red marker identifies the target position as given in the TIC, while concentric circles represent the photometric aperture and sky annulus used in the flux extraction (see Section~\ref{sec:photometry}).
The animation is particularly useful to assess whether the apparent variability originates from the target itself or from nearby variable stars or moving sources such as asteroids.

Figure \ref{fig:cutout} illustrates this diagnostic for the young eclipsing binary Brun 691, using a 20×20-pixel TESS cutout. The frame-by-frame animation, available on the GitHub pages\footnote{Frame-by-frame animation: \url{https://javiserna.github.io/TESSExtractor-webapp/assets/frame_by_frame.gif}} shows that the measured variability originates exclusively from the central source, no nearby objects within the TESS field exhibit correlated brightness changes over time. The photometric aperture and sky annulus remain well centred throughout the cadence sequence, and no background transients, asteroids, or pointing-related drifts are apparent in the animation. This provides a direct visual confirmation that the LC extracted for Brun 691 is not affected by time-dependent contamination from nearby stars.

\subsection{Pixel by pixel LCs}
In addition to the integrated aperture photometry, the advanced option of TESSExtractor allows the computation of pixel-by-pixel LCs across the full cutout region.
After normalising each pixel’s flux to its median value, the software arranges the resulting time series into a $10\times10$, $15\times15$ or $20\times20$ mosaic of miniature panels.
Each panel corresponds to an individual detector pixel, plotted on the same time axis as the global LC; coherent variability across adjacent panels indicates that those pixels are dominated by a genuine stellar signal, whereas uncorrelated or highly noisy panels indicate contamination from nearby objects.

The right panel of Figure \ref{fig:cutout} shows the pixel-by-pixel LC mosaic for Brun 691. Despite the moderate crowding in this field (CM = 18\%), the spatial distribution of variability is highly localised: the central 3×3 pixels display nearly identical modulation, while all surrounding pixels show flat or uncorrelated behaviour. No peripheral pixel exhibits similar periodic or eclipse-like signatures that would suggest astrophysical contamination from neighbouring sources within the 20×20 pixel cutout. This confirms that the dominant variability arises from Brun 691; neighbouring blended sources primarily contribute static flux dilution and do not show coherent time-variable signatures in the pixel-by-pixel light curves. Therefore, this diagnostic complements the animation in Section \ref{sec:frame_by_frame} by demonstrating that the extracted LC is astrophysically clean, even in moderately crowded fields.

\subsection{Systematic Effects Correction}
\label{sec:systematics}
In the systematic-correction mode, we first normalise the LC flux by its median value and then use the mission-produced Cotrending Basis Vectors (CBVs), a set of orthonormal vectors that describe the dominant sources of variability shared among multiple stars in the TESS cameras but not intrinsic to the stellar signal \citep{SPOC_2016, Caldwell_2020}. Typical effects include CCD-related systematics and scattered light from the Earth and Moon. Up to 16 CBVs may be provided, numbered from 1 to 16 in order of decreasing importance, though the number of populated vectors can vary by sector, camera, and CCD \citep{Caldwell_2020}. We use the first four CBVs, which carry the strongest and best-defined systematic trends with low noise components; including higher-order CBVs in an unconstrained least-squares fit risks overfitting, thereby inadvertently removing intrinsic astrophysical variability from the LC \citep{Smith_2012}.

\begin{figure*}
    \centering
    \includegraphics[width=0.7\textwidth]{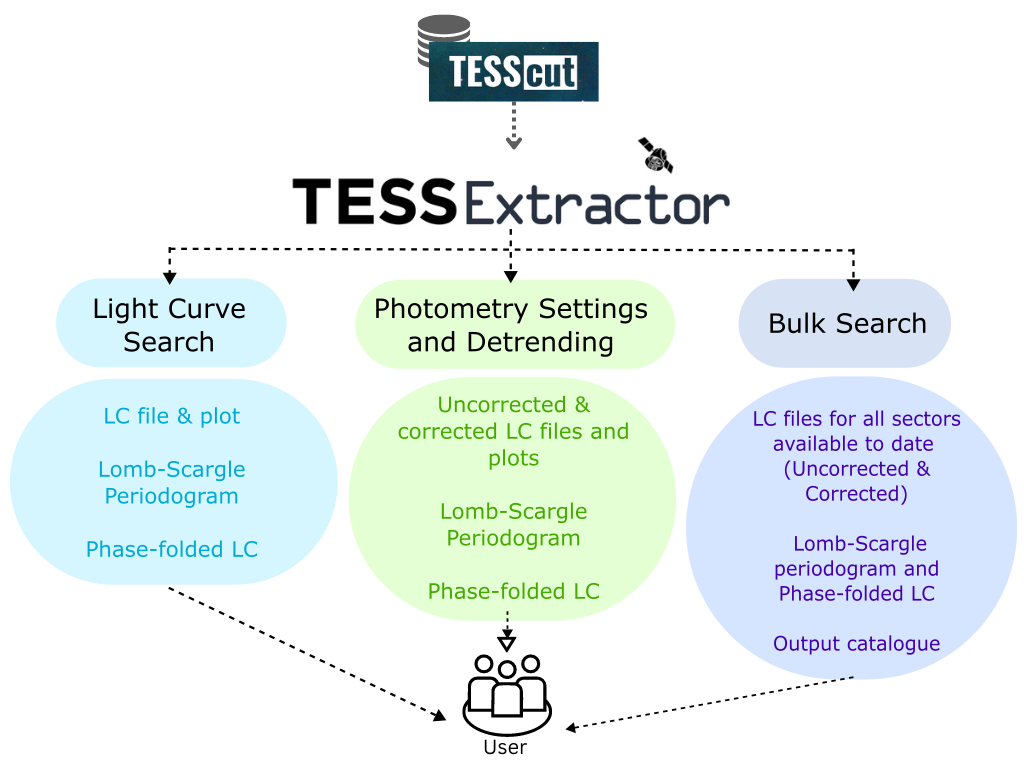}
    \caption{Flowchart diagram of the TESSExtractor web application. Rounded labels denote the three processing modes, coloured circles group their respective data products, and dashed arrows indicate the data flow from \texttt{TESScut} through each mode to the user.}
    \label{fig:flowchart}
\end{figure*}

To correct the normalised LC for systematic effects, we built the representative systematics vector as the linear combination of the CBVs, and then obtain the systematics-corrected flux, which is expressed as:
\begin{equation}
    F_{corr}=F_{\ast}-F_{sys}\equiv F_{\ast}-\sum_{i=1}^{N}a_{i}\times CBV_{i}
\end{equation}
where $F_{\ast}$ is the normalised flux, $F_{sys}$ is the representative systematics vector, and the index ($i$) goes from 1 to 4, and $a_{i}$ is the scaling coefficient, given by the least-squares solution as $a_{i}=(A^{T}A)^{-1}A^{T}F_{\ast}$, where A is the normal matrix with CBVs components as columns.
We apply a $\chi^{2}$-test, where $\chi^{2}=\sum_{j}{(F_{\ast,j}-F_{sys,j})^{2}}/{(F_{\ast,j})}$, with the \textit{j} index running over the data points of the LC. The correction is made until $\chi^{2}$ is minimised or the process reaches 100 tests. Figure \ref{fig:sys_corr} illustrates the systematic correction process. The bulk mode applies two independent corrections per sector, a singlescale (SS) correction using the first four vectors, and a multiscale+spike (MS) correction using the first vector from each of multiscale bands 1-3 and the spike layer (four vectors total), both fitted via iterative robust least-squares with $5\sigma$ clipping. The uncorrected LC, the SS-corrected flux, and the MS-corrected flux are all delivered for each target and sector.

\subsection{Deployment, Accessibility, and User Modes}
TESSExtractor is developed in Python and implemented using the Streamlit framework, an open-source library for building web applications\footnote{\includegraphics[width=0.15in, height=0.15in]{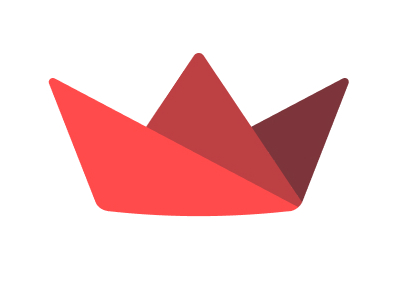}\href{https://streamlit.io/}{https://streamlit.io/}}. Our web application is hosted on a cloud service, running on a dedicated machine that operates continuously 24/7 with on-demand network bandwidth and memory, allowing simultaneous access by multiple users. Additionally, our bulk search service is hosted on our institutional virtual machine at the University of Oklahoma.\\

\begin{figure*}
    \centering
    \includegraphics[width=0.35\textwidth]{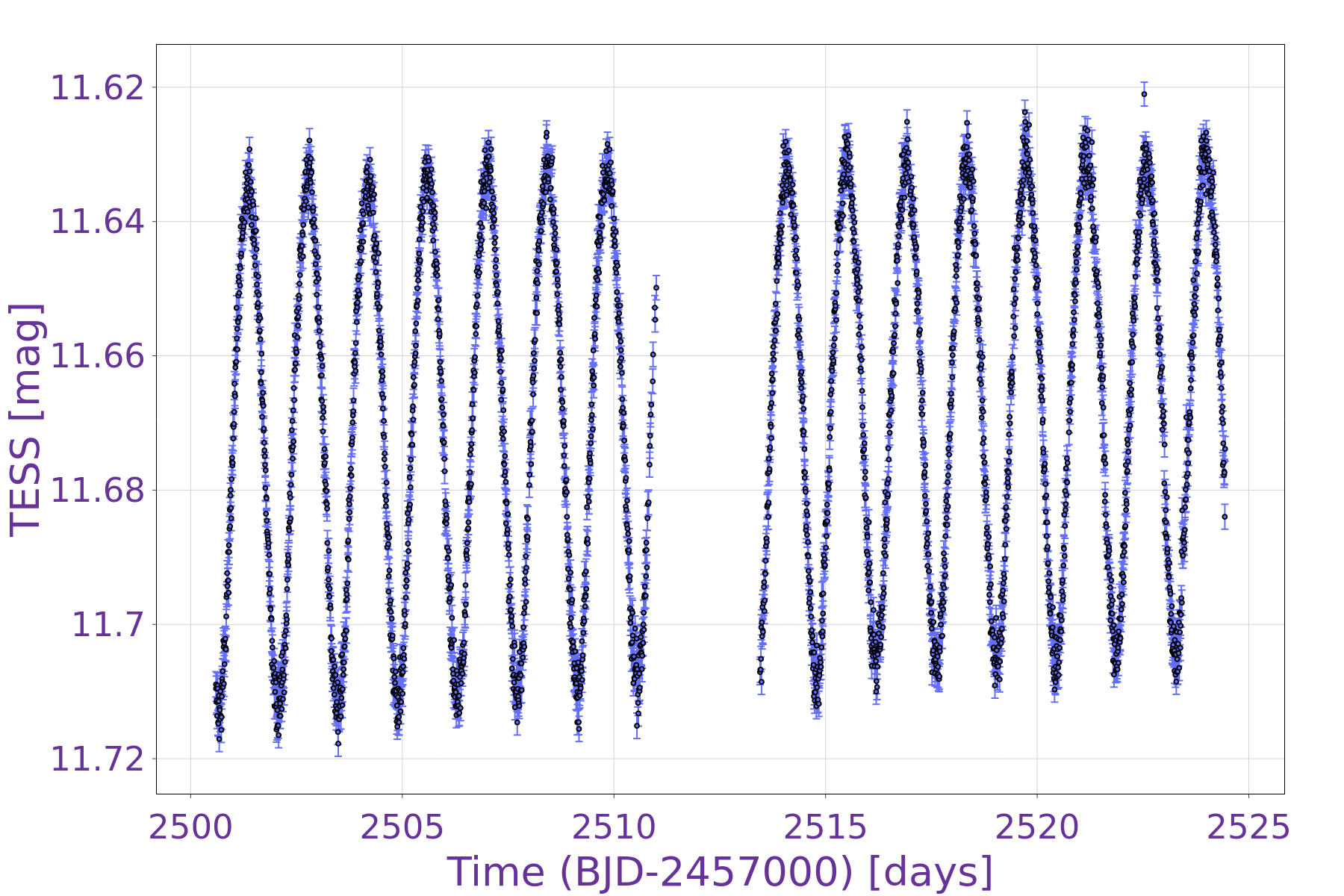}
    \includegraphics[width=0.25\textwidth]{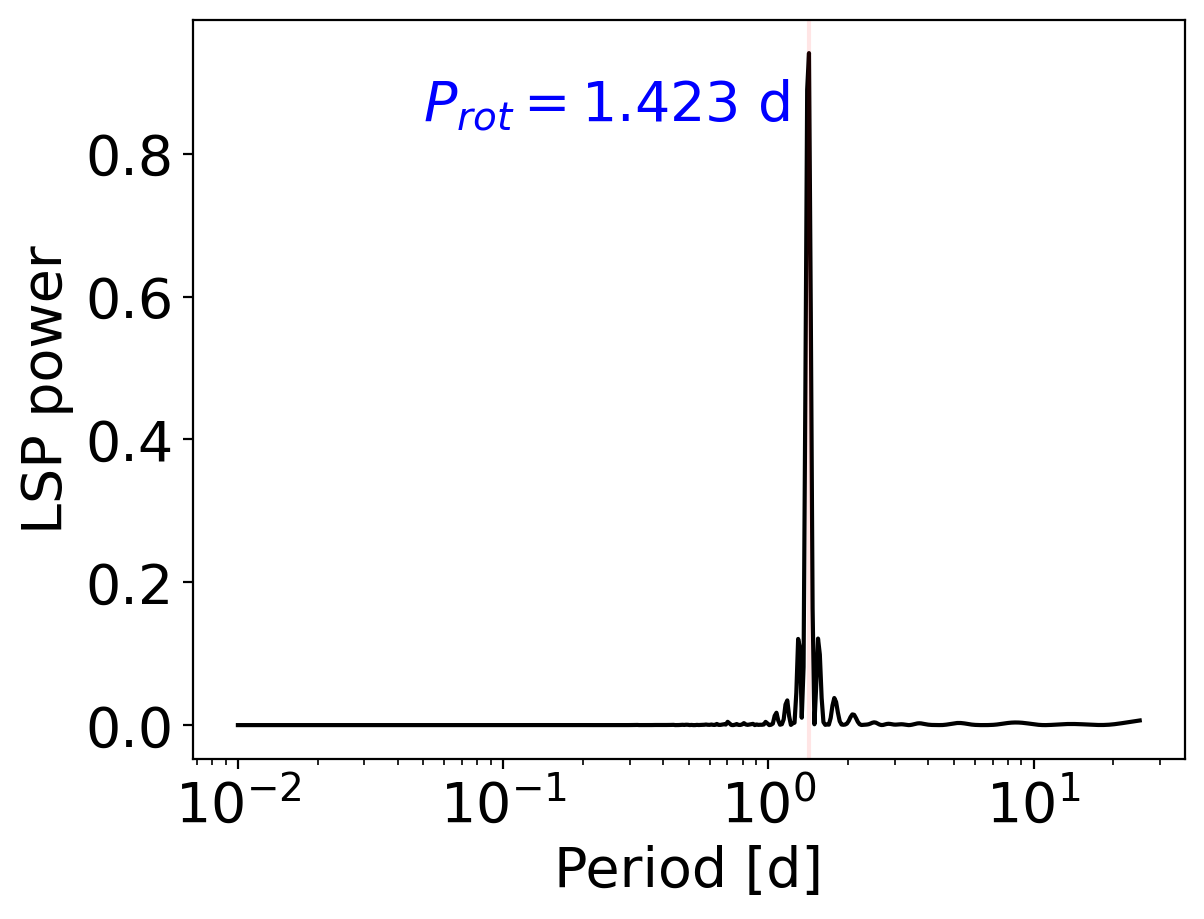}
    \includegraphics[width=0.265\textwidth]{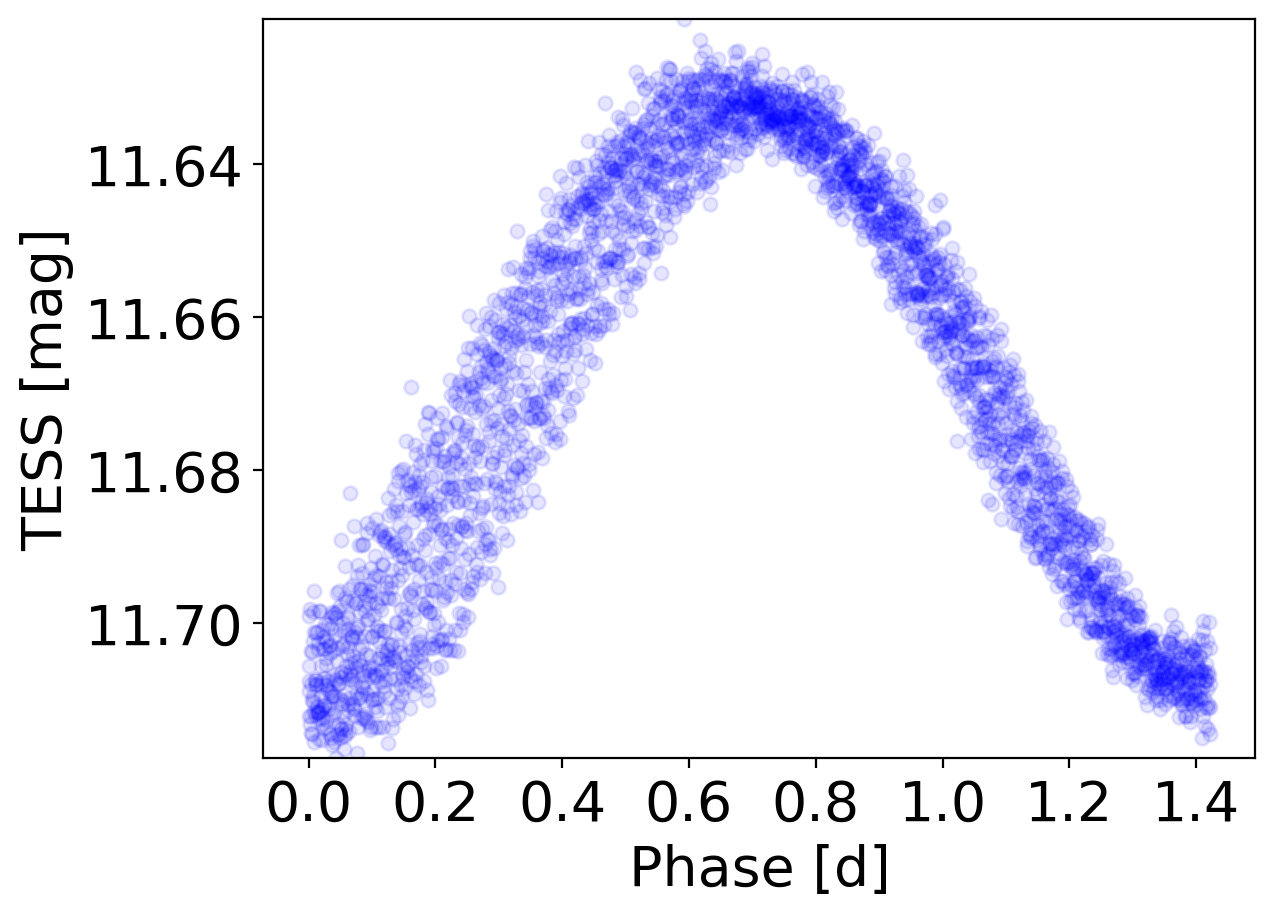}
    \caption{Primary data product of TESSExtractor for TIC 58229237, LC, periodogram, and phase-folded LC.}
    \label{fig:dataproduct}
\end{figure*}

In the default and systematic correction options, the user inputs the name or coordinates of the target, target names 
are normalised to their canonical catalogue form and resolved against TIC~v8.2 \citep{Stassun_2019} via \texttt{astroquery} 
\citep{Ginsburg_2019}, returning the nearest TIC source within the search radius; raw J2000 coordinates follow the same 
cross-match. The resolved TIC position, identifier, and $T_{\rm mag}$ are then used throughout the pipeline. After clicking on the search button, and the app automatically begins to download FFI cutouts, process the LCs, and displays the corresponding data products (LC, periodogram, phase-folded LC, DSS and TESS cutouts). Users can also download the corresponding LC and periodogram in CSV data format. The \textit{Bulk Search} option allows users to upload a list of targets. The uploaded file must be in plain ASCII text, with target names or coordinates listed one per line. Once the list is successfully uploaded, the user's email is requested, and then a request button is displayed. This option sends the input targets to the job queue, and users are then notified via email that their request has been scheduled.
When our institutional virtual machine detects new user requests in the queue, it proceeds to download TESS data, processes all available LCs per TESS sector, and generates and compresses all resulting products into a ".zip" file, which is subsequently sent to the user by email. For further information about the operability, user modes, and data products,  please refer to Figure \ref{fig:flowchart}.

\subsection{Data products}

\subsubsection{LC, Periodogram, and Phase LC plots}
The main product of our application is the LC, obtained following the procedure of Section~\ref{sec:photometry}.
To enable users to explore and analyse these data more deeply, we use Plotly \citep{plotly} to display an interactive scatter plot of the TESS magnitude (equation \ref{eq:mag}) as a function of the time. We also incorporate error bars estimated using equation \ref{eq:magerr}.
Our interactive visualisation offers features such as zooming, panning, and hover-over tools, which facilitate efficient data exploration. The plot can be downloaded in ".png" format for sharing or offline use.

Upon processing of the LC, we utilise the Lomb-Scargle periodogram \citep{Lomb_1976,Scargle_1982} to search for periodic signals in the LC. We sample 500 trial periods, uniformly spaced in $\log P_{\rm rot}$ between 0.01 and 25 days. The best period is selected as the one corresponding to the highest peak in the periodogram. The choice of 500 trial periods balances period resolution ($\Delta P/P \lesssim 1.6\%$ across the sampled range) against computational cost, ensuring near-real-time response for interactive use. To assess the statistical significance of the detected period, we compute the false alarm probability (FAP) of the highest periodogram peak using the analytic approximation of \citet{Baluev_2008}, as implemented in \texttt{Astropy} \citep{astropy:2022}. The FAP quantifies the probability that a peak of the observed power or higher would arise from Gaussian noise alone; lower values therefore indicate more reliable detections. The FAP value of the best period is displayed numerically on the periodogram, together with three horizontal significance levels at FAP $= 10\%$, $1\%$, and $0.1\%$.
Simultaneously a phase-folded LC is generated. Using the task \texttt{FoldAt} of PyAstronomy \citep{pya}, the LC is folded at the best selected period. Users may also specify a custom folding period through a dedicated input field; in that case, the LC is refolded at the user-provided value instead of the periodogram peak. An example of each data product is shown in Figure \ref{fig:dataproduct}.

\subsubsection{File Format}
In addition to the visual and interactive products, we provide users with access to downloadable data files that contain the information needed to reproduce the displayed plots.\\

Data files are available for all three modes. For the \textit{default} mode, two CSV files are produced per sector: the LC file (\texttt{\{target\}\_s\_\{sector\}.csv}) with columns \texttt{time}, \texttt{mag}, and \texttt{mag\_err}, and the periodogram file (\texttt{\{target\}\_per\_s\_\{sector\}.csv}) with columns \texttt{period} and \texttt{power}. Users can download these files via the ``Download'' buttons on the portal.

For the \textit{systematic correction} mode, two separate LC files are produced: \texttt{\{target\}\_s\_\{sector\}\_uncorrected.csv} and \texttt{\{target\}\_s\_\{sector\}\_corrected.csv}, both with columns 
\texttt{time}, \texttt{flux}, and \texttt{flux\_err}. The periodogram file (\texttt{\{target\}\_s\_\{sector\}\_periodogram.csv}) contains columns \texttt{period} and \texttt{LSpower}.

For the \textit{bulk search} mode, data are delivered as plain-text files with seven columns per cadence: \texttt{Time}, \texttt{Tmag}, \texttt{Tmag\_err}, \texttt{Flux\_CBV\_SS}, \texttt{Flux\_err\_SS}, \texttt{Flux\_CBV\_MS}, and \texttt{Flux\_err\_MS}, providing the 
uncorrected SAP magnitude alongside both SingleScale- and 
MultiScale+Spike-corrected fluxes for each target and sector.

To optimise cloud resources, we automatically refresh the cache every 3 minutes, and the final data product remains available on the website for this duration or until the user initiates another search.

\section{Validation and Applications}
\label{sec:Applications}
\subsection{LC comparison}

To evaluate the photometric performance of \texttt{TESSExtractor} relative to existing TESS pipelines, we performed a direct comparison between FFI-extracted LCs produced by our tool and those generated by the SPOC, QLP, \texttt{Eleanor}, and TGLC pipelines, as publicly available through the MAST archive. For each pipeline, we retrieved the corresponding LCs for the same target and processed them in a homogeneous manner to ensure a fair comparison.

\begin{figure*}
    \centering
    \includegraphics[width=1\textwidth]{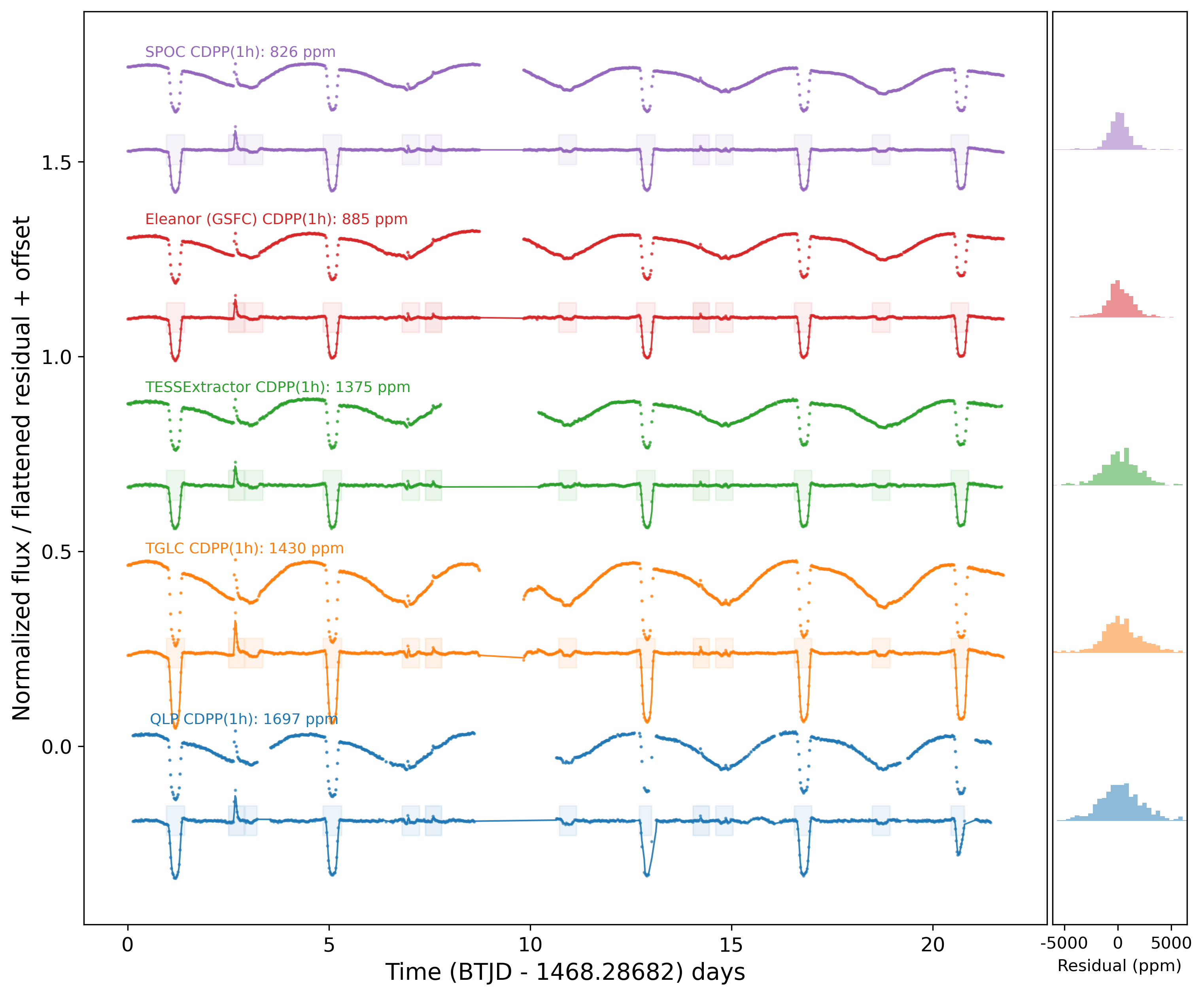}
    \caption{Comparison of TESS LCs for the young binary Brun~691 (Sector 6) extracted with five different pipelines: the SPOC pipeline (violet), \texttt{Eleanor} \citep{Feinstein_2019} (red), \texttt{TESSExtractor} (green), TESS--Gaia LCs \citep{Han_2023} (TGLC; orange), and MIT’s Quick-Look Pipeline \citep{Huang_2020} (QLP; blue).  All LCs are shown on a common time axis, vertically offset for clarity, and detrended using the same flattening procedure. Shaded regions mark intervals masked prior to the CDPP calculation (e.g., eclipses and flare events). For each pipeline, we annotate the 1-hour CDPP, which quantifies the photometric precision and allows a direct comparison of noise properties across methods. The right panels show the distribution of flattened residuals without transits and flares, illustrating differences in scatter and noise symmetry among pipelines.}
    \label{fig:validation}
\end{figure*}

All LCs were detrended using the \texttt{Lightkurve} \texttt{.flatten()} routine \citep{Lightkurve_2018}, which removes long-term trends using a Savitzky--Golay filter, with a window length of 51 cadences (polyorder = 2) to remove long-term trends, allowing us to assess short-timescale photometric precision on equal footing across pipelines. The resulting LCs were normalised and vertically offset for visual clarity in Figure~\ref{fig:validation}.

As a quantitative metric, we computed the Combined Differential Photometric Precision (CDPP), originally defined for the \textit{Kepler} mission as a measure of the effective white-noise level on transit-relevant timescales \citep{Christiansen_2012}. Conceptually, CDPP corresponds to the depth of a box-shaped transit signal that would be detected with a signal-to-noise ratio of unity over a given integration time; lower CDPP values therefore indicate higher photometric precision. In this work, we estimate CDPP on 1-hour timescales for all pipelines.

To isolate the impact of low-level noise and variability, we compute CDPP on a cleaned version of each LC in which eclipses and strong astrophysical signals (such as flares) are masked prior to the calculation; these intervals are identified visually from the detrended LC as regions where the flux deviates by more than $3\sigma$ from the running baseline, and the shaded regions in Figure~\ref{fig:validation} mark the masked intervals. This approach focuses the metric on the residual low-amplitude fluctuations after removing large-amplitude signals; these fluctuations may reflect the instrumental noise floor, photometric scatter, or intrinsic low-level stellar variability such as granulation, or residual accretion variability.

Figure~\ref{fig:validation} shows the comparison for the test case Brun~691, a young eclipsing binary composed of a solar-type star and a brown dwarf \citep{Morales-Calderon_2012}. The LC is extracted from Sector 6, which has a 30-minute cadence. All pipelines recover the dominant astrophysical variability and eclipse morphology consistently, demonstrating that durations, depths, and overall timing are preserved independently of the extraction method.

For this target, the contamination metric computed by our pipeline (Section~\ref{sec:contamination}) is CM = $18\%$, indicating a moderate level of flux dilution from neighbouring Gaia sources; for comparison, the SPOC pipeline reports \texttt{CROWDSAP}=0.893 for the same target, corresponding to $10.7$\% contamination (1-\texttt{CROWDSAP}).

While the Gaia-based CM ($18\%$) nominally exceeds our recommended $15\%$ caution threshold (Section~\ref{sec:contamination}), this is consistent with CM being a conservative, Gaia-anchored estimate relative to PRF-based metrics in crowded fields (Section~\ref{sec:contamination}). The SPOC-based \texttt{CROWDSAP} value ($10.7\%$) falls below the $20\%$ threshold (crowding metric $\geq$ 0.8) adopted by \citet{Caldwell_2020} to select TESS-SPOC field-star targets with reliable aperture photometry, and the overall variability and eclipse morphology remain robust across all apertures.

At the same time, small but measurable differences in the scatter around the baseline are apparent at the sub-percent level. These differences reflect the distinct methodologies adopted by each pipeline. \texttt{Eleanor} minimises CDPP on hour-long timescales to optimise sensitivity to planetary transits, while SPOC first performs aperture photometry using SNR-optimised apertures to produce \texttt{SAP\_FLUX}, and subsequently applies CBVs derived from ensemble systematics to remove common instrumental trends, yielding the \texttt{PDCSAP\_FLUX} product. QLP relied on fixed apertures optimised for bright stars, whereas TGLC employs PSF-based photometry informed by Gaia, which enables deblending of sources in crowded fields, effectively mitigating flux contamination from neighbouring stars. We note that, as of Sector 94, QLP has adopted the TGLC photometry method \citep{Petitpas_2026}, a transition that postdates the Cycle 2 data used in this comparison.

These distinctions account for the differences in residual scatter visible in the right panels of Figure~\ref{fig:validation}; the broader distributions for \texttt{TESSExtractor}, TGLC, and QLP are consistent with their shared 30-minute FFI cadence, while \texttt{Eleanor}'s narrower distribution reflects its CDPP-minimising aperture weighting, and SPOC's narrower distribution partly reflects its correspondingly shorter physical detrending window (51 cadences $\times\,2\,\mathrm{min} \approx 1.7\,\mathrm{h}$ versus $\sim$25\,h for 30-minute pipelines). These differences therefore reflect cadence and optimisation strategy rather than intrinsic photometric quality.

In addition to differences in the noise properties, we observe a systematic variation in the eclipse depths recovered by the different pipelines. In particular, TGLC produces noticeably deeper primary and secondary eclipses compared to the aperture-based pipelines (SPOC, \texttt{Eleanor}, QLP, and our own \texttt{TESSExtractor}). This behaviour is expected and arises from the PSF-based forward modelling adopted by TGLC, which uses Gaia positions and fluxes to model and subtract the contribution of all neighbouring sources within the TESS cutout. By mitigating flux dilution from blended stars, TGLC isolates a cleaner estimate of the target flux, naturally yielding deeper eclipse profiles. Aperture-based methods, even with optimised radii and background subtraction, inevitably include some contaminating flux due to the 21 arcsec TESS pixel scale, resulting in shallower measured eclipse depths \citep{Han_2023}. The consistency among SPOC, \texttt{Eleanor}, and \texttt{TESSExtractor} therefore reflects comparable levels of aperture dilution, while the increased depth in TGLC highlights the advantages of PSF-based deblending in moderately crowded fields.

The LC extracted with \texttt{TESSExtractor} exhibits a photometric precision and baseline stability that are comparable to those achieved by \texttt{Eleanor} and SPOC, as reflected in the CDPP estimates. In several segments, the flattened residuals show slightly reduced low-frequency structure, likely due to the combination of local background estimation and catalogue-position-anchored aperture selection implemented in our pipeline. Importantly, all genuine astrophysical signals, including eclipses and rotational modulations, are consistently detected across all datasets, although their amplitudes differ systematically between aperture and PSF based methods, as discussed above.

\begin{figure}
    \centering
    \includegraphics[width=\columnwidth]{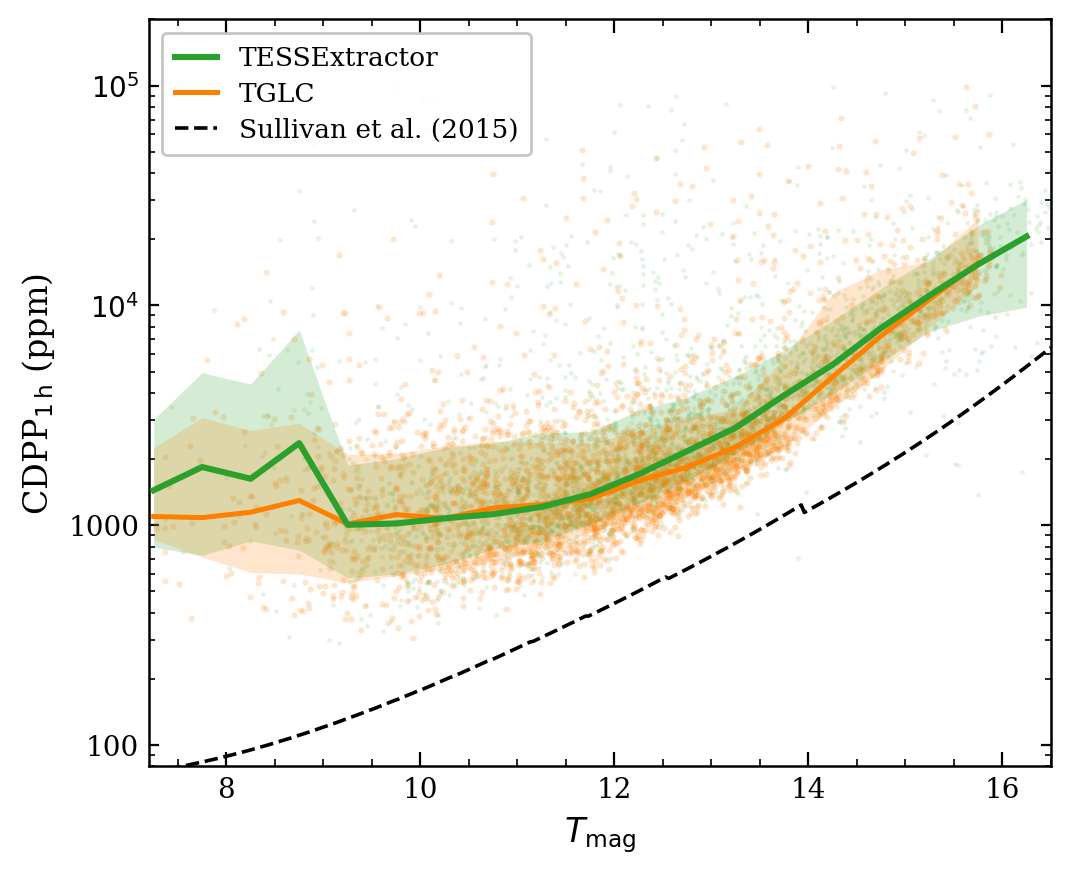}
    \caption{CDPP$_{1\,\rm h}$ as a function of TESS magnitude for \texttt{TESSExtractor} (green; $N=34{,}088$) and TGLC \citep[orange;][]{Han_2023} ($N=7{,}150$), both operating on TESS FFIs from the young stellar sample of \citet{Kounkel_2019b}.
    Solid lines show the running median in 0.5\,mag bins; shaded regions span the 16th--84th percentile dispersion of each distribution. Sources with CDPP$_{1\,\rm h}>10^{5}$\,ppm are excluded as spurious. The dashed line shows the theoretical photon-noise floor from \citet{Sullivan_2015}.}
    \label{fig:cdpp_comparison}
\end{figure}

Overall, this single-target comparison demonstrates that \texttt{TESSExtractor} recovers the correct LC morphology and CDPP-level precision on a representative young binary. To assess whether this performance is consistent across the full TESS magnitude range, we extend the comparison statistically in Figure~\ref{fig:cdpp_comparison}.

To extend the photometric comparison beyond a single test case, Figure~\ref{fig:cdpp_comparison} shows CDPP$_{1\,\rm h}$ as a function of $T_{\rm mag}$ for \texttt{TESSExtractor} and TGLC across a sample of $\sim$34{,}088 and $\sim$7{,}150 light curves, respectively, drawn from the young stellar population of \citet{Kounkel_2019b}. Sources with CDPP$_{1\,\rm h} > 10^{5}$\,ppm were excluded as spurious. We compare specifically against TGLC because both pipelines operate directly on TESS FFIs and cover the same magnitude range down to $T_{\rm mag}\approx16$, making them the most directly comparable products in terms of cadence and input data.

Across the full magnitude range, the median CDPP$_{1\,\rm h}$ of \texttt{TESSExtractor} tracks that of TGLC within the 16th--84th percentile dispersion of both distributions, indicating broadly consistent photometric performance at the population level. At $T_{\rm mag}\gtrsim11$, where the majority of \texttt{TESSExtractor} use cases reside, the two pipelines are essentially equivalent. At the bright end ($T_{\rm mag}\lesssim10$), TGLC shows a lower running median, though the two distributions overlap substantially in this regime and the small number of sources precludes a definitive comparison.
Both pipelines lie above the theoretical photon-noise floor of \citet{Sullivan_2015}, as expected for a young stellar population that retains residual low-amplitude variability after detrending.

\subsection{Rotational Period}
\label{sec:period}

\begin{figure*}
    \centering
    \includegraphics[width=1\textwidth]{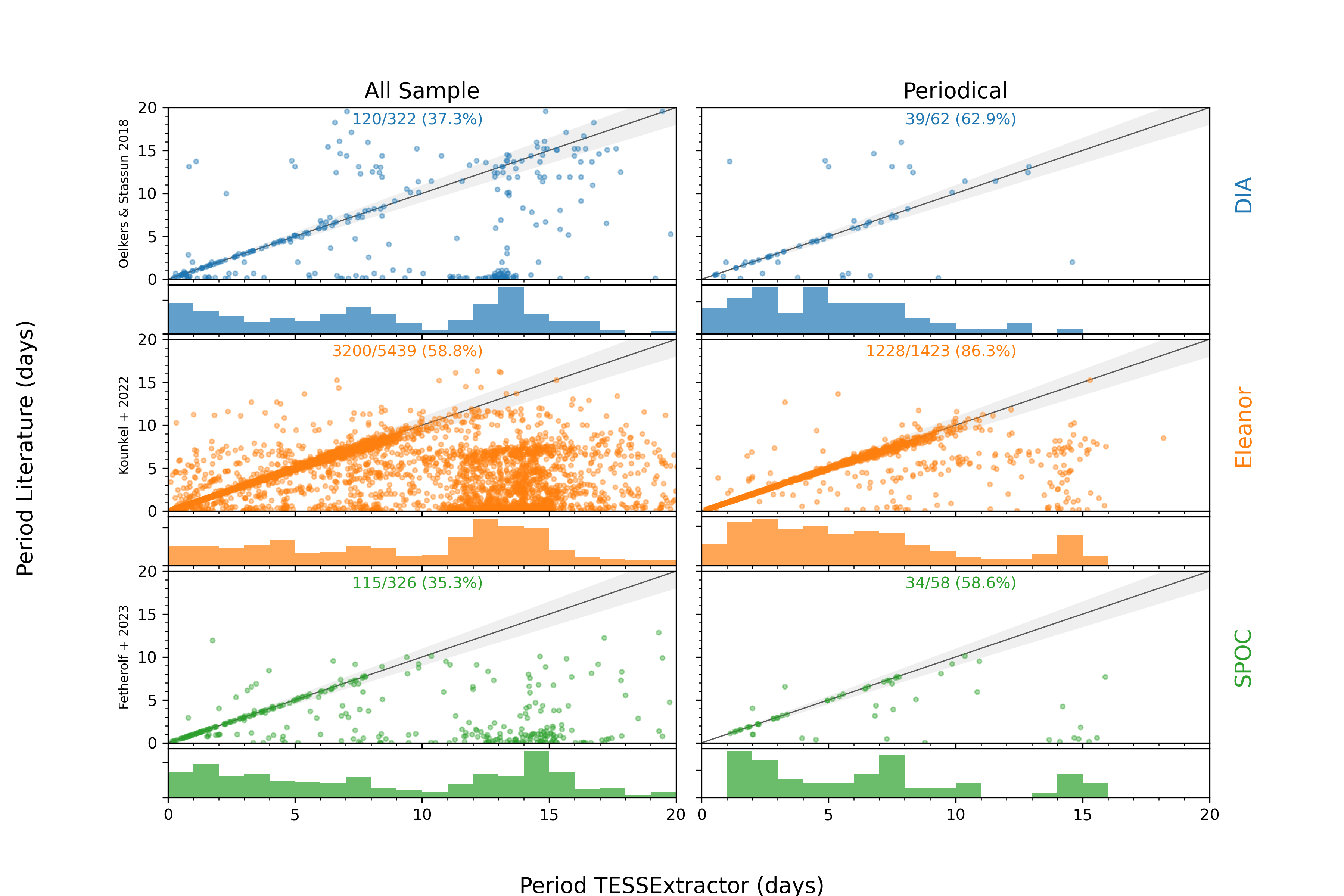}
    \caption{Comparison between rotation periods derived with \texttt{TESSExtractor} and literature values from three independent catalogues: DIA-based periods from \citet{Oelkers_2018} (top panels), \texttt{Eleanor} periods from \citet{Kounkel_2022} (middle panels), and SPOC periods from \citet{Fetherolf_2023} (bottom panels). The left column shows the full cross-matched samples, while the right column displays only the LCs classified as periodic by the morphology-based ML classifier (Section~\ref{sec:classification}). In each panel, the fraction of stars consistent with the 1:1 relation within a 10\% tolerance (i.e.\ $0.9\,P_{\rm TESSExtractor}<P_{\rm Literature}<1.1\,P_{\rm TESSExtractor}$) is shown in the corresponding catalogue colour. Histograms beneath each scatter plot show the distribution of \texttt{TESSExtractor} periods for matched sources. Across all catalogues, agreement fractions increase substantially for the periodic subset (right column), demonstrating that morphological classification efficiently removes non-periodic or systematics-dominated LCs and improves period recovery by 23-29 percentage points relative to the unfiltered sample.}
    \label{fig:comp1}
\end{figure*}

In this subsection, we assess the reliability of rotation periods derived with \texttt{TESSExtractor} by comparing them directly with independent literature catalogues based on TESS FFIs. Our primary reference is the young, clustered stellar population identified by \citet{Kounkel_2019b}, for which \citet{Kounkel_2022} measured rotation periods using \texttt{Eleanor} \citep{Feinstein_2019} LCs and Lomb-Scargle analysis. This population spans ages from $\sim1$ to $100$ Myr and includes a heterogeneous mixture of periodic, quasi-periodic, stochastic, and systematics-dominated LCs.

Not all TESS LCs are expected to yield a stellar rotation period. Young stars frequently exhibit accretion bursts, extinction dips, multi-periodic signals, and evolving spot patterns that violate the assumptions of classical periodogram searches \citep{Hedges_2018, Cody2018}. In addition, instrumental systematics in TESS FFIs introduce spurious power at characteristic timescales, most notably near $\sim12$ days \citep{Kounkel_2022}. As a consequence, a strict one-to-one correspondence between catalogues is not anticipated when comparing all LCs indiscriminately.

We extracted $34,088$ LCs from the \citet{Kounkel_2019b} catalogue using \texttt{TESSExtractor}. Rotation periods were searched on the light curves as extracted by TESSExtractor, without applying CBV corrections, to avoid suppressing low-frequency astrophysical variability on timescales comparable to the rotation periods probed here. Cross-matching with \citet{Kounkel_2022} yielded $16,936$ sources in common, of which $5,439$ have reported rotation periods in both datasets. We define period agreement conservatively as  
\[
0.9\,P_{\rm TESSExtractor} \;<\; P_{\rm Literature} \;<\; 1.1\,P_{\rm TESSExtractor},
\]  
which accounts for finite periodogram resolution.

Using the full, heterogeneous sample, we find that $3,200/5,439$ sources ($58.8\%$) satisfy this criterion (Figure~\ref{fig:comp1}). The remaining discrepancies primarily correspond to stars with irregular or ambiguous variability that is unlikely to produce a single, well-defined period that can be associated with stellar rotation.

To isolate a more reliable subset, we applied supervised machine-learning classifiers trained on the morphological taxonomy defined by \citet{Elizabethson_2023} (see Section \ref{sec:classification} for a full 
description of the classification framework). These models are not used to measure periods, but exclusively to identify LCs that exhibit stable periodic variability consistent with stellar rotation. The classifiers, originally trained and validated on young stars in Orion, have demonstrated high reliability in distinguishing periodic from stochastic behaviour or other types of variability.

Restricting the comparison to the $1,423$ LCs classified as periodic, we find that $1,228$ satisfy the agreement criterion, corresponding to an $86.3\%$ recovery fraction. This sharp improvement indicates that discrepancies in the full sample arise primarily from intrinsically non-periodic or systematics-dominated LCs, rather than from limitations in the \texttt{TESSExtractor} period determination itself.

We extend this validation using fully independent methodologies. Cross-matching with the Difference Image Analysis (DIA) pipeline of \citet{Oelkers_2018}, accessed via the Filtergraph portal,\footnote{\url{https://filtergraph.com/tess_ffi}} confirms consistent period distributions across overlapping TESS sectors (2, 3, 4, 5, 15, 16, and 17; Figure~\ref{fig:comp1}). We also compare against rotation periods from the TESS prime mission catalogue of \citet{Fetherolf_2023}. For sources in common, the agreement fractions are $120/322$ ($37.3\%$) for Oelkers \& Stassun and $115/326$ (35.3\%) for Fetherolf et al.\ in the full sample, increasing to $62.9\%$ and $58.6\%$ for periodic LCs; these fractions are fully consistent with the trends observed in the Kounkel et al.\ comparison (Figure~\ref{fig:comp1}).

Taken together, these results demonstrate that \texttt{TESSExtractor} recovers rotation periods consistent with independent catalogues when applied to LCs exhibiting clear periodic morphology. Disagreement with literature values in the full sample predominantly reflects intrinsic stellar variability or data systematics rather than methodological failure. When morphology-aware filtering is applied, \texttt{TESSExtractor} yields rotation periods in good agreement with independent analyses based on \texttt{Eleanor}, DIA, and SPOC LCs, with recovery fractions of 86.3\%, 62.9\%, and 58.6\%, respectively. This validates the use of \texttt{TESSExtractor} as a reliable tool for stellar rotation studies, particularly in young and heterogeneous stellar populations.

\subsection{Light Curve Morphology Classification}
\label{sec:classification}

Understanding the morphology of a LC is essential for interpreting its underlying physical mechanism, particularly in young stellar objects where multiple processes such as rotation, accretion, extinction, pulsations, and binarity, may coexist. While rotation period measurements (Section~\ref{sec:period}) provide a quantitative characterization of periodic variability, they do not capture differences in waveform shape or amplitude evolution, which are often key diagnostics of accretion, disk-related phenomena, transiting events or multiplicity.\\

To complement our period validation and assess the broader scientific utility of the extracted LCs, we applied the machine-learning classification scheme developed by \citet{Elizabethson_2023}. Their model defines eleven morphological classes commonly observed in T Tauri stars, including Periodic (P), Multiperiodic (MP), Dipper/Burster (DB), Eclipsing Binary (EB), Long-term variability (L), and Noisy (N). \citet{Elizabethson_2023} also included other periodic morphological class variants, such as Periodic+Noise (Pn), Periodic+an overall increasing brightness (Pi), Periodic+an overall decreasing brightness (Pd), LCs with increasing amplitude
in the periodic signal (Pia), and LCs with decreasing amplitude in
the periodic signal (Pda).

These classes were originally calibrated using high-quality TESS FFI-based LCs of young stars, the same data type produced by \texttt{TESSExtractor} and capture the dominant signatures associated with rotation, accretion bursts, extinction dips, and stochastic variability. This provides a consistent observational basis for applying the classification scheme to the LCs extracted by TESSExtractor.

\begin{figure*}
    \centering
    \includegraphics[width=1\textwidth]{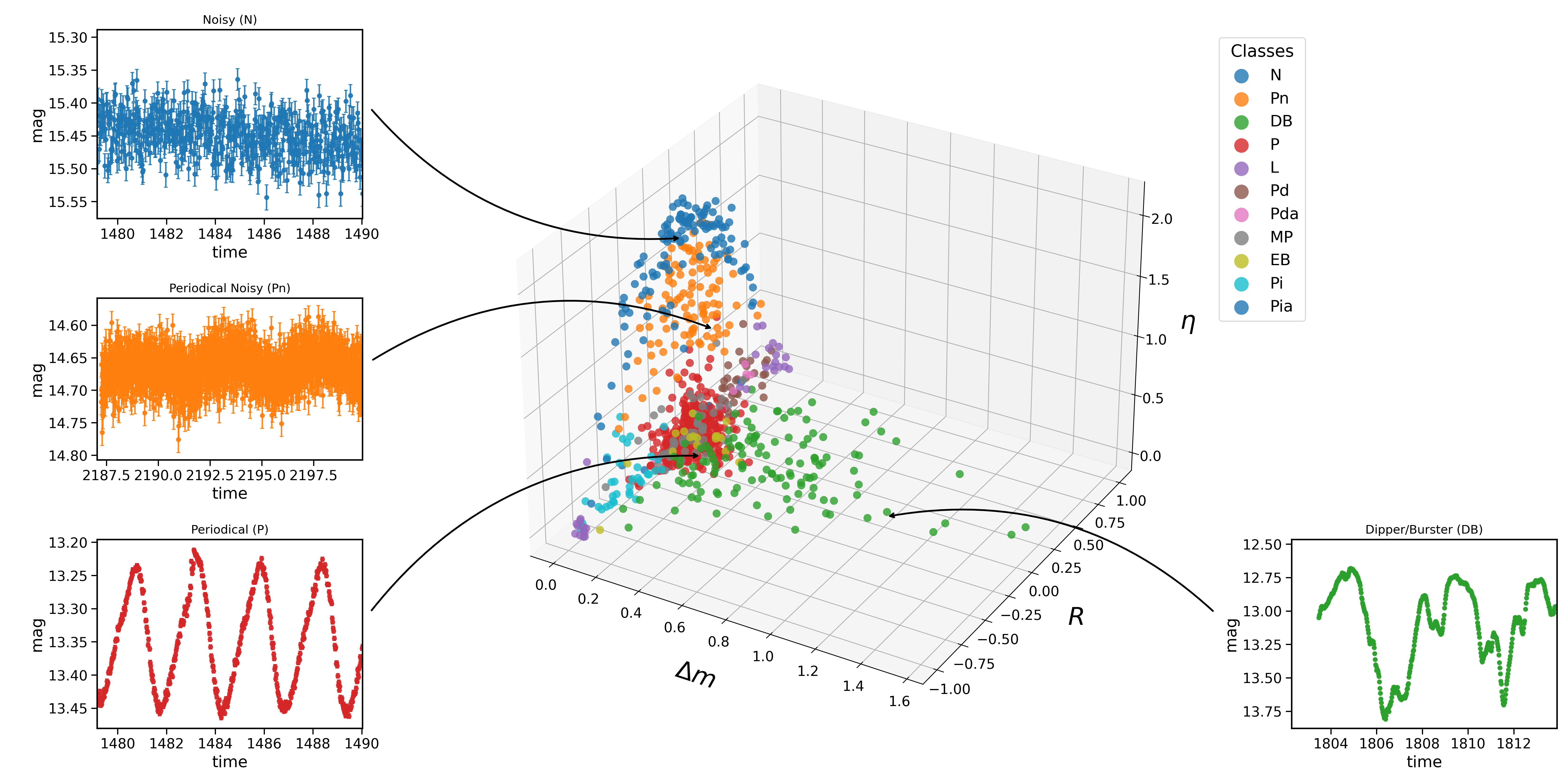}
    \caption{Example of light-curve morphological classes from \citet{Elizabethson_2023} applied to a subset of TESSExtractor LCs drawn from the \citet{Kounkel_2019b} catalogue. The central 3D feature space illustrates how three diagnostic metrics such as $\Delta m$ (the amplitude range of the LC), $R$ (the linear correlation coefficient), and $\eta$ (the von Neumann ratio, a proxy for short-timescale noise or stochasticity), naturally cluster according to variability-morphological classes. Representative LCs are shown around the diagram for several classes, including Noisy (N), Periodical Noisy (Pn), Periodical (P), and Dipper/Burster (DB), demonstrating how distinct morphological behaviours map onto well-separated regions of this reduced feature space. This visualisation highlights the discriminative power of even a small subset of 3 out of 28 features used in the machine-learning classifier of \citet{Elizabethson_2023}.}
    \label{fig:sources}
\end{figure*}

Each LC is characterised by the set of 28 variability metrics defined by \citet{Elizabethson_2023}(Table~\ref{tab:lc_features_example}), encoding statistical, temporal, and periodogram-based properties; following their feature-space reduction (their Section~3.3), the pre-trained classifier uses the 18 most informative of these features. To demonstrate how the extracted features map into observable variability patterns, Figure \ref{fig:sources} provides a visual example of how different morphological classes occupy distinct regions of the three dimensional feature space.

Although the full feature set is multidimensional, we illustrate in Figure \ref{fig:sources} how even a subset of three features can already differentiate several morphological groups, highlighting the discriminative power of the methodology.

We computed the full set of 28 LC metrics for each source extracted by \texttt{TESSExtractor}, and applied the pre-trained classifier using the corresponding 18 features to assign morphological labels across our sample (Table \ref{tab:lc_features_example}). This automated classification enables two key outcomes:

\begin{enumerate}
    \item Filtering of period measurements:  
    By isolating only the morphologies known to yield reliable periods (e.g., P and Pn), we increase the agreement with the literature (Section~\ref{sec:period}), demonstrating the importance of morphology-aware period validation.

    \item Identification of non-periodic or contaminated LCs: 
    Classes such as DB, N, or L frequently exhibit poor period recovery due to systematics or intrinsic irregularity, and the classifier provides an efficient way to flag such cases.
\end{enumerate}

The feature-extraction code will be made publicly available in a future release, ensuring reproducibility and facilitating incorporation into other FFI-processing pipelines. In addition to the rotation-period validation, the ML-based classification demonstrates that \texttt{TESSExtractor} produces LCs of sufficient quality for both quantitative and morphological time-domain studies.

\begin{table*}
\centering
\caption{LC features computed for the morphology classifier (Section~\ref{sec:classification}). The table lists basic identifiers, the full set of 28 variability metrics, and the predicted morphological label. This table is published in its entirety in the online supplementary material; a portion is shown here for guidance regarding its form and content.}
\label{tab:lc_features_example}
\scriptsize
\setlength{\tabcolsep}{2pt}
\renewcommand{\arraystretch}{1.05}
\resizebox{\textwidth}{!}{%
\begin{tabular}{llllllllllllllll}
\toprule
RA & DEC & TIC & mad & os & low & row & fm & sm & tm & tail & per & per\_2 & per\_3 & slope & r\_value \\
\midrule
310.6077 & 30.6548 & 100088837 & 0.0095 & -0.0288 & -0.1350 & 0.2171 & 0.5660 & 0.2940 & 0.1130 & 0.0200 & 14.1860 & 6.8180 & 2.8472 & -0.0002 & -0.0992 \\
71.2997 & 54.8516 & 10012676  & 0.0050 &  0.1224 &  0.0369 & 0.0206 & 0.1050 & 0.0640 & 0.0250 & 0.5140 & 12.0130 & 5.2450 & 3.0720 &  0.0007 &  0.7539 \\
71.3546 & 55.0831 & 10013094  & 0.0008 & -0.0419 & -0.1525 & 0.1276 & 0.0530 & 0.0350 & 0.0320 & 0.2060 &  4.3210 & 12.0880 & 3.0720 & -0.0001 & -0.4821 \\
71.3745 & 55.1734 & 10013258  & 0.0011 & -0.0543 & -0.1165 & 0.1378 & 0.0560 & 0.0380 & 0.0330 & 0.3810 & 11.9380 &  3.0720 & 4.2458 & -0.0001 & -0.6397 \\
71.3836 & 55.1951 & 10013290  & 0.0026 & -0.0394 & -0.0973 & 0.2229 & 0.1590 & 0.0850 & 0.0370 & 0.1740 & 12.0880 &  2.9720 & 3.4716 &  0.0002 &  0.3664 \\
\bottomrule
\end{tabular}%
}
\vspace{4pt}
\resizebox{\textwidth}{!}{%
\begin{tabular}{llllllllllllllll}
\toprule
slope\_min & r\_value\_min & eta & DeltaM & dtw & Env & USlope & LSlope & med & rV & reucliD & reDSign & rbLeon & rbLeonSign & robAbbe & Classes \\
\midrule
 0.0000 &  0.0071 & 0.7332 & 0.0544 & 4.1952 & -0.0006 & -0.0008 & -0.0002 & 13.4280 & 0.3381 & 0.0220 & -0.0208 & 0.0062 & -0.0001 & 1.9573 & Pn+N \\
 0.0007 &  0.7573 & 0.4073 & 0.0178 & 3.3867 & -0.0008 &  0.0007 &  0.0015 & 13.4130 & 0.1426 & 0.0211 & -0.0208 & 0.0026 &  0.0001 & 1.9749 & N+L  \\
-0.0005 & -0.8418 & 1.1588 & 0.0045 & 0.9049 & -0.0001 & -0.0001 & -0.0001 & 10.5690 & 0.0407 & 0.0209 & -0.0208 & 0.0008 & -0.0000 & 2.0237 & Pia+N \\
 0.0000 &  0.0001 & 0.8155 & 0.0054 & 1.0278 & -0.0001 & -0.0002 & -0.0001 & 11.2160 & 0.0455 & 0.0209 & -0.0208 & 0.0009 & -0.0000 & 1.9101 & N    \\
-0.0002 & -0.4142 & 0.9738 & 0.0138 & 2.5054 &  0.0001 &  0.0002 &  0.0001 & 13.0890 & 0.1245 & 0.0210 &  0.0208 & 0.0022 & -0.0000 & 1.8886 & N    \\
\bottomrule
\end{tabular}%
}
\end{table*}

\section{Discussion}
\label{sec:Discussion}
\subsection{Limitations and future developments}

\subsubsection{User experience}
The graphical interface follows a straightforward linear workflow that guides users from target selection to data visualisation in a vertically ordered sequence. This layout enhances clarity and accessibility, particularly for non-expert users, by presenting each processing step in a logical and self-contained manner. However, the cumulative output of interactive plots and animations can progressively saturate the workspace, requiring increased vertical scrolling as additional products are rendered. The Streamlit architecture currently lacks support for advanced window management features, such as tabbed layouts or dockable panels, which would allow users to collapse, rearrange, or customise interface components according to their preference. Consequently, while the linear design promotes transparency and ease of navigation, it limits spatial efficiency and user-level customisation within the browser environment. To mitigate these constraints and preserve overall usability, advanced diagnostics such as the pixel-by-pixel LCs and frame-by-frame cutout animation, are excluded from the default workflow. These products can nevertheless be enabled, allowing users to visualise all products simultaneously.

\subsubsection{Concurrent users}
Because Streamlit executes Python code synchronously in a single event loop, the current implementation can handle only a limited number of simultaneous sessions. Empirically, we find that a single server instance can support roughly five concurrent users before the latency associated with downloading FFIs, performing aperture photometry and generating periodograms becomes noticeable. Requests for multiple sectors or large pixel cutouts are particularly resource intensive and can block the event loop for several seconds. This limitation is more evident in sectors with short cadence (e.g., <10 minutes). To improve scalability we are developing a worker–pool architecture in which computationally expensive tasks (data download, photometry, periodogram calculation) are dispatched to asynchronous worker processes. Horizontal scaling deploying multiple replicas and load-balancing sessions across them, will further mitigate concurrency limitations and isolate user sessions. Under this architecture each user interacts with an independent worker, avoiding race conditions and ensuring consistent performance even under high traffic conditions.

\subsubsection{Crosstalk between sessions}
During the early stages of development, we occasionally observed session crosstalk, in which cached pixel data were inadvertently shared between simultaneous user sessions under heavy load. This issue was traced to globally scoped cache objects that persisted across sessions. The problem has since been resolved by implementing per-session cache identifiers and automatic cache clearing whenever a new target is queried. Although no crosstalk has been observed in recent deployments, future versions will further strengthen session isolation by sandboxing cache operations within dedicated worker namespaces, ensuring robust separation between concurrent users.

\subsubsection{Browser compatibilities}
The web interface works across desktop, laptop, and even mobile devices, and relies on modern browser technologies such as WebGL and the \texttt{Plotly.js} library to render interactive plots and frame-by-frame animations. Performance and compatibility may vary among browsers: the application runs optimally on Google Chrome, Mozilla Firefox, and Microsoft Edge, which natively support WebGL~2.0 and hardware acceleration. Safari, particularly on older macOS versions, may require manual activation of the “WebGL via Metal” setting to display animations properly. When WebGL support is unavailable, the system automatically falls back to static image rendering to preserve usability. Users are advised to employ the latest desktop versions of Chrome or Firefox to ensure full interactivity. A real-time browser compatibility check notifies users when interactive features are unavailable, although all photometric processing and data products remain fully functional.

\subsubsection{Cloud resources}
At the infrastructure level, the application is hosted on Railway, which currently provides 30 virtual CPUs and 32 GB of RAM. These resources have proven sufficient for current usage patterns, and the system has not yet reached performance saturation. Nonetheless, both vertical and horizontal scaling options are available, including vertical scaling by expanding the allocated memory and CPU, and horizontal scaling by replicating app instances, allowing \texttt{TESSExtractor} to accommodate future growth in user demand without compromising responsiveness.

\subsubsection{Input coordinates versus identifiers}
A recurrent source of uncertainty arises when users query targets by coordinates rather than by identifiers. Because the application internally cross-matches with the TIC v8.2 \citep{Stassun_2019} to locate nearby sources, coordinate-based queries can sometimes resolve to neighbouring TIC objects in crowded regions. To ensure unambiguous source identification, it is recommended that users provide \textsc{SIMBAD}-resolved names or TIC identifiers whenever possible. Future versions may include automated disambiguation tools that visually display all nearby TIC sources within the query radius, helping users confirm the target of interest.

\subsubsection{Queue system}
The \textit{Bulk Search} mode employs a first-in, first-out queue to handle multi-target requests. Each job corresponds to a list of targets that are processed sequentially across all available sectors. While this approach guarantees fairness and reproducibility, targets located in the TESS continuous viewing zone (CVZ) can significantly increase processing time, as they are observed in numerous consecutive sectors. Planned developments include job prioritization, sector-based task sharing, and the provision of estimated completion times to improve throughput and transparency. Parallel execution of individual sectors across multiple worker nodes is expected to reduce overall latency for both CVZ and non-CVZ targets.

\subsubsection{CCD borders}
Targets situated near TESS CCD edges may experience partial aperture truncation or, when the sky annulus extends beyond the cutout boundary, a reduction in the number of valid pixels available for background estimation. No explicit proximity buffer is implemented to flag or exclude such sources; instead, \texttt{TESSExtractor} relies on the native pixel-masking behaviour of \texttt{Photutils}, which restricts the background computation to the unmasked pixels within the annulus. When the valid annular area is significantly reduced, the background estimate becomes less reliable, which is reflected as elevated 
photometric uncertainties in the LC. Users are therefore advised to inspect the uncertainty bars carefully when analysing sources near CCD boundaries, as these serve as the primary diagnostic for edge-related degradation. The implementation of automatic edge-proximity checks and user-facing warning flags is planned for a future release.

\subsubsection{DSS2 versus TESS Cutout}
After rotating the TESS image to match the DSS2 orientation (north up and east to the left), we found that, in some cases, the central position of the TESS cutout appears offset by approximately one TESS pixel relative to the DSS2 frame. This small shift arises from the intrinsic resolution of the TESS WCS solution, which is limited by the 21 arcsec pixel scale and by interpolation effects introduced during the rotation and reprojection process. Although this offset does not affect the photometric extraction, it can produce a slight apparent mismatch between the TESS and DSS2 centroids when the two fields are visually compared.

\subsubsection{Saturation and contamination}
The quality of the extracted photometry depends strongly on the target’s brightness and the local background environment. Stars fainter than $T_\mathrm{mag}\approx 16$ have photometric uncertainties exceeding 5\%, reducing the utility of their LCs for high-precision studies. Conversely, stars brighter than $T_\mathrm{mag}\approx 6.8$ may saturate the TESS detector \citep{Sullivan_2015}, leading to non-linear flux measurements and potential contamination of neighbouring pixels. In crowded regions near the Galactic plane or in star-forming clusters, flux from multiple sources can blend within a single photometric aperture, complicating background subtraction and the determination of reliable periods. We therefore recommend using caution when analysing targets at the extremes of the TESS dynamic range or in densely populated fields. The incorporation of additional crowding metrics and automatic warning flags in the application will be part of future development, particularly to quantify the impact of the extended wings of bright stellar PSFs on nearby pixels.

\subsubsection{Aperture-corrected magnitude calibration}
The magnitudes reported by \texttt{TESSExtractor} in default and bulk modes ($m^{\ast}$; Section~\ref{sec:photometry}) are not absolute photometric measurements. They result from anchoring the median instrumental magnitude to the catalogued $T_{\rm mag}$ from TIC~v8.2 \citep{Stassun_2019}, which may differ from the star's true brightness during the observed sector. Users working with intrinsically variable sources, including eruptive young stellar objects, long-term variables, or eclipsing systems, should treat the reported magnitudes with caution, as the correction may introduce a systematic offset relative to the true astrophysical brightness.

\subsubsection{Magnitude uncertainty approximation}
Equation~\ref{eq:magerr} is a first-order approximation obtained by linearising $m = -2.5\log_{10} F_{\ast}$, which yields symmetric error bars of the form $\delta m \approx 1.085\,\delta F / F_{\ast}$. This approximation incurs a relative error of $\approx 1/(2\,\mathrm{SNR})$ in the upper magnitude uncertainty. For sources with $\mathrm{SNR} \gtrsim 10$ ($\delta F / F_{\ast} \lesssim 10\%$), this error remains below $5\%$, which is acceptable for the visualisation purposes of \texttt{TESSExtractor}. At lower SNR, the magnitude uncertainties become intrinsically asymmetric: the upper error (towards fainter magnitudes) exceeds the lower one, with exact values given by $\sigma_m^{+} = -2.5\log_{10}(1 - \delta F/F_{\ast})$ and $\sigma_m^{-} = 2.5\log_{10}(1 + \delta F/F_{\ast})$. Reporting asymmetric error bars will be incorporated in a future release.

\subsubsection{Measuring variability between TESS sectors}
\texttt{TESSExtractor} applies a normalisation process to the LC using a catalogued Tmag that is independent of the TESS sector. Although \texttt{TESSExtractor} LCs are well-suited for visualising and analysing brightness variability within an individual TESS sector or observing epoch, they cannot currently be used for inter-sector brightness-variation studies with the current implementation. To assess long-term brightness variations across multiple TESS sectors, it is necessary to use unnormalised aperture fluxes and then apply a sector-by-sector relative calibration, for example, using non-variable field stars observed in the same sector. At present, inter-sector brightness variability is beyond the scope of our pipeline.

\subsection{Support and Development}
This application supports a broad range of time-domain investigations, including stellar rotation, activity monitoring, eclipsing binaries, planetary transits, and asteroseismology and is particularly well suited for rapid visualisation and inspection of TESS LCs. The application is continuously maintained, upgraded and monitored to ensure stable performance, incorporate new functionalities and provide an efficient and reliable service to the astronomical community.

\section{Conclusions}
We have presented \texttt{TESSExtractor}, an open-source web-based application that allows astronomers and the broader community to extract, visualise and analyse TESS LCs directly from Full-Frame Images. By integrating aperture photometry, cotrending corrections, Lomb–Scargle periodogram analysis and interactive visualisations within a Streamlit-based interface, \texttt{TESSExtractor} democratises access to TESS time-domain data. We validated the photometric performance of \texttt{TESSExtractor} through a morphological comparison against SPOC, QLP, \texttt{Eleanor}, and TGLC for a representative young binary (Brun~691), and through a statistical comparison of CDPP$_{1\,\rm h}$ versus $T_{\rm mag}$ against TGLC across $\sim$34{,}088 FFI-extracted light curves, demonstrating consistent photometric performance at the population level. We also applied the tool to tens of thousands of young stars to measure stellar rotation periods and used machine-learning models to classify LC variability morphologies. Limitations of the current implementation, including concurrent user handling, browser compatibility and queue latency, were discussed together with planned technical improvements. Future developments include expanded multi-sector processing capabilities, improved cache and session management, and the incorporation of additional interactive pixel-level tools. \texttt{TESSExtractor} is publicly available at \url{https://www.tessextractor.app} and is designed to evolve continuously through community feedback and scientific use.

\section*{Acknowledgements}
We thank Allison Youngblood for facilitating connections with members of the TESS Science Support Center that contributed to this work. The authors thank Nicole Schanche for the helpful discussions and comments that have helped to improve this manuscript and the app. J.S. thanks Jérôme Bouvier for assistance with the orientation of TESS images relative to DSS frames, and Nuria Calvet for testing the application and providing comments on an early version of the manuscript. J.S. also acknowledges support from the Homer L. Dodge Department of Physics and Astronomy at the University of Oklahoma. J.H. acknowledges support from the UNAM-DGAPA-PAPIIT research projects IG-101723, and IN110126. R. L-V. acknowledges support from Secretar\'ia de Ciencia, Humanidades, Tecnolog\'ia e Innovaci\'on (SECIHTI) through a postdoctoral fellowship within the program ``Estancias posdoctorales por M\'exico''.
We thank the Activity and Rotation of Young Stellar Objects (ARYSO) collaboration for their valuable discussions and contributions.

This paper includes data collected with the TESS mission, obtained from the MAST data archive at the Space Telescope Science Institute (STScI). Funding for the TESS mission is provided by the NASA Explorer Program.

This research made use of \texttt{Photutils}, \texttt{Lightkurve}, and \texttt{Astropy} packages for detection and photometry of astronomical sources.

\section*{Data Availability}

All TESS FFI data used in this work are publicly available from the MAST archive at https://archive.stsci.edu/tess. Processed light-curve products demonstrated in this paper can be reproduced through the public web interface at https://www.tessextractor.app.



\bibliographystyle{rasti}
\bibliography{references_short_version}







\label{lastpage}
\end{document}